The Effect of Macromolecular Crowding, Ionic Strength and Calcium Binding on Calmodulin Dynamics

Qian Wang[*], Kao-Chen Liang[*], Arkadiusz Czader[&], M. Neal Waxham[+] and Margaret S. Cheung[*#]
*Department of Physics, University of Houston, Houston, TX 77204
&Department of Chemistry, University of Houston, Houston, TX 77204
+Department of Neurobiology and Anatomy, University of Texas, Health Science Center, Houston, TX
# Corresponding author: mscheung@uh.edu

Short title: *Calmodulin Dynamics under Cell-like Conditions*

**Abstract:** The flexibility in the structure of calmodulin (CaM) allows its binding to over 300 target proteins in the cell. To investigate the structure-function relationship of CaM, we combined methods of computer simulation and experiments based on circular dichroism (CD) to investigate the structural characteristics of CaM that influence its target recognition in crowded cell-like conditions. We developed a unique multiscale solution of charges computed from quantum chemistry, together with protein reconstruction, coarse-grained molecular simulations, and statistical physics, to represent the charge distribution in the transition from apoCaM to holoCaM upon calcium binding. Computationally, we found that increased levels of macromolecular crowding, in addition to calcium binding and ionic strength typical of that found inside cells, can impact the conformation, helicity and the EF hand orientation of CaM. Because EF hand orientation impacts the affinity of calcium binding and the specificity of CaM's target selection, our results may provide unique insight into understanding the promiscuous behavior of calmodulin in target selection inside cells.

## Author Summary

Proteins are workhorses for driving biological functions inside cells. Calmodulin (CaM) is a protein that can carry cellular signals by triggered conformational changes due to calcium binding that alters target binding. Interestingly, CaM is able to bind over 300 targets. One of the challenges in characterizing CaM's ability to bind multiple targets lies in that CaM is a flexible protein and its structure is easily modulated by the physicochemical changes in its surroundings, particular inside a complex cellular milieu. In order to determine structure-function relationships of CaM, we employed a combined approach of experiments, computer simulations and statistical physics in the investigation of the effect of calcium-binding, salt concentration, and macromolecular crowding on CaM. The results revealed unique folding energy landscapes of CaM in the absence and presence of calcium ions and the structural implications of CaM are interpreted under cell-like conditions. Further, a large conformational change in CaM in response to environmental impacts, dictates the packing of local helices that may be critical to its function of target binding and recognition among vast target selections.

**Introduction**

Calmodulin (CaM) is a highly acidic protein composed of 148 amino acid residues [1,2]. It has four helix-loop-helix calcium binding motifs commonly named EF-hands (Figure 1). Two EF-hands form a lobe and the two lobes (termed N and C) are connected by a helical linker. The linker has hinge like characteristics that accounts for extensive interdomain movements. One of the most remarkable characteristics of CaM is its ability to bind over 300 targets[3], owing to its conformational flexibility. Without calcium binding, the crystal structure of CaM was in an extended, dumbbell structure [4]. The EF-hand orientation is antiparallel and the hydrophobic residues are buried inside the protein[4]. However, after calcium binding, CaM becomes more compact [5,6] where the N- and C-lobes collapse and the orientation of EF hands is perpendicular[4,7]. Through such conformational change, the hydrophobic residues, especially Met [8], are exposed to the solvent and this facilitates CaM binding to targets. Although this $Ca^{2+}$-induced conformational transition is crucial for the binding ability of CaM to a subset of targets [9], there are a number of proteins that can bind to the calcium-free form of CaM[10,11]. As shown in other anisotropic proteins with the ability to change shapes [12,13,14], CaM structures are susceptible to macromolecular crowding effects [15,16] under which the most probable ensemble structure of CaM is a collapsed form in the absence of calcium[17]. The structural response of the N- and C-lobes to the level of crowding is different and this likely plays an important part in influencing CaM's promiscuous binding behavior. These implications suggested a complicated behavior of CaM as a signaling protein in a cell. However, our understanding of CaM conformational dynamics under conditions that more closely mimic those inside the cell is still unknown.

Protein modeling and computer simulations can provide critical insights into CaM's structural flexibility. Although all-atomistic molecular dynamics simulations can reveal the structures of CaM in great detail, they are often constrained by the short simulation times accessible (e.g. nanoseconds) so they cannot fully address the conformational changes that take place on the longer time scale relevant to calcium and target binding [18,19,20]. To overcome this hurdle, coarse-grained molecular simulations (CGMS) [17,21,22,23] were developed to probe the plastic structure of CaM. CGMS is particularly efficient for studying protein dynamics under the effect of macromolecular crowding[24] and this method has been used to study protein folding mechanism under cell-like conditions [25,26,27,28]. However, the interactions between coarse-grained residues may miss important ingredients that account for the functionality of CaM. For example, pairwise Lennard-Jones potentials are deployed to represent the London dispersion forces, assuming that the electrostatic interactions are coarse-grained into a parameter in solvent-mediated interactions [29]. However, this approach misses critical information related to electrostatic interactions among charged residues, particularly when they are close in space. Another approach many groups have adopted is a so called "double Go-type" interaction, but such approaches can only sample the differences between two protein states based on *a priori* known structures [21,22]. The latter approach misses the opportunity to explore other possible compact structures in solutions [30,31]. Although the electrostatic interactions between the charged amino acids or nucleic acids have been investigated by coarse-grained model [32,33], there are few combined experimental and computational studies that attempt to validate or confirm the predictions of the computational work [34,35].

Motivated by the need for a better description of CaM conformational dynamics under cell-like conditions, we introduced a coarse-grained protein model that takes into account the electrostatic interaction between charged residues under different ionic strength conditions through the Debye- Hückel potential [36]. The main challenge lies on the determination of charges on CaM that are dependent on calcium binding and altered ionic strength and the resulting extent of solvent exposure in such a structure. We overcome this issue by incorporating a new approach that combines reconstruction of an ensemble of all-atomistic protein models from coarse-grained simulations (MultiSCAAL) [37], statistical physics and quantum chemistry. With this new method, the charges distribution of apoCaM and holoCaM are reasonably computed and the effects of electrostatic interactions as a function of ionic strength on CaM, as well as the level of macromolecular crowding, can be evaluated.

We have probed the structural characteristics of apoCaM in response to changes in ionic strength and macromolecular crowding effects. At high ionic strength, an extended dumbbell-like state of apoCaM is most probable and its helicity increases. At high crowding level, a collapsed compact state of apoCaM is most probable and the alignment of the two helices of an EF-hand was found to change, exposing buried hydrophobic residues. Although these two environmental factors can both result in an increased absorption of far-UV CD signals, different types of compact apoCaM structures may cause the change in the signal. This work reinforces the growing appreciation for the need to examine proteins in environments that mimic the intracellular milieu if one desires to understand their interactions and functions inside cells.

# Results

## Results from experiments

*Modulation of CaM's conformation and thermal stability by calcium binding*

CaM is a highly dynamic molecule whose conformational states in solution have been probed through the use of CD. As a prelude to investigate the impact of ionic strength and macromolecular crowding, we first analyzed the far-UV CD spectra and thermal unfolding of CaM in the absence and presence of $Ca^{2+}$. In the apo-state (Figure 2A), CaM displays negative bands at 207 and 222 nm consistent with the CD spectrum of a molecule with a relatively high proportion of α-helical content. In the presence of saturating $Ca^{2+}$, there was a consistent and significant increase in the negative bands at both 207 and 222 nm. While some have interpreted this change as an increase in α-helical content, others have suggested that this difference in the spectrum is plausibly due to reorientation/distortions of the existing α-helices [38]. The latter suggestion is consistent with NMR data comparing apo- and holoCaM [7] showing that the α-helical content of the two states are similar. We additionally examined the thermal stability of CaM by monitoring the CD spectra at 222 nm (Figure 2B) while increasing the temperature. Under these conditions of low ionic strength, apoCaM is completely unfolded at 60ºC and above. The data also indicate that multiple transitions exist from the folded to the unfolded state (described in more detail below). As reported previously [39], addition of $Ca^{2+}$ leads to a stabilization of the structure of CaM where even at temperatures of 90ºC, the protein retains evidence of a partially folded state (Figure 2B).

*Ionic strength impacts the far-UV CD spectra and thermal denaturation of apoCaM*

Ionic strength influences a variety of protein bonding potentials. To investigate how ionic strength influences the conformation of CaM, we evaluated the CD spectrum of apoCaM at increasing KCl concentrations. As seen in Figure 3A, step-wise increases in KCl between 0 and 250 mM, produced systematic but modest increases in the negative bands at 208 and 222 nm. A further increase to 500 mM KCl did not produce additional increases in the CD signal. The thermal stability of CaM at increasing ionic strength was also assessed. Figure 3B shows averaged traces of thermal denaturation data of bulk (0 added KCl) up to 500 mM KCl. Visual inspection of the data also revealed that there was a non-monotonic shape in the melting curve suggesting the presence of at least one intermediate in the pathway from the folded to the unfolded states. To more quantitatively examine these transitions, the data in Figure 3B were fit to a three-state model. Fitting to a two-state model produced unacceptably large residuals in some of the data, so a three-state model including an intermediate state (I) between the native folded (N) and the unfolded (U) states was employed. The use of a three-state model was also employed in an earlier report examining the thermal stability of apoCaM where the transitions were proposed to reflect the unfolding of first the C-terminal lobe, followed by the N-terminal lobe[40]. Parameter values from models fits to this data, $T_m$ and $\Delta H_m$, are provided in Table 1. Increasing ionic strength significantly increased the $T_m$ of both the C- and N-lobes of CaM (N-I and I-U, respectively). Although the errors from the fitting for the van't Hoff $\Delta H$ were large in some cases, there was a slight trend for increasing $\Delta H$ for the C-lobe (N-I), but no significant trend for the N-lobe (I-U). These data indicate that ionic bonding potential plays a significant role in stabilizing apoCaM's structure, however, the impact of $Ca^{2+}$ binding, even in the absence of KCl, produces an even greater resistance to thermal denaturation (Figure 2 and Table 1).

*Modulation of apoCaM's structure and thermal unfolding by macromolecular crowding*

To investigate how macromolecular crowding influences the structure of CaM, the CD spectra of apoCaM was determined at increasing concentrations of the crowding agent Ficoll 70. Ficoll 70 is a non-interacting, non-ionic glucose polymer that has been used extensively to investigate crowding induced effects on protein structure and folding. As seen in Figure 3C, increased concentrations of Ficoll 70 (from 100 – 400 mg/ml) produced systematic increases in the CD spectra of apoCaM. Note that there was a subtle change in the structure from 0 – 100 mg/ml, then a more significant impact at 200 mg/ml and then modest additional impact on the spectra by increasing the concentration to 300 or 400 mg/ml.

We additionally assessed thermal unfolding of apoCaM by monitoring changes in CD at 222 nm in the presence of increasing amounts of Ficoll 70. Figure 3D shows that increasing concentrations of Ficoll 70 (0 – 400 mg/ml) produced a systematic shift to the right in the melting temperature of apoCaM. Fits of this data to a 3-state model were accomplished and the parameters from those fits are presented in Table 2. Quantitatively, the thermal midpoint temperature of the N-I transition for apoCaM increased from 24.3 to 48.4°C while the I-U transition increased from 51.8 to 65.9°C after the addition of 40% Ficoll 70.

We additionally examined the combined impact of crowding and ionic strength on the thermal unfolding of apoCaM. Figure 3, panels E and F display the data and parameters derived from fits to the data are shown in Table 2. As noted previously (Figure 3 and Table 1), inclusion of salt has a significant impact on the thermal stability of apoCaM in the absence of crowding agents. In these experiments, 100 mM KCl produced an increased thermal midpoint temperature, $T_m$, for the N-I and I-U transitions from 24.3 to 46.2 and from 51.8 to 61.4°C, respectively. The van't Hoff enthalpy at $T_m$, $\Delta H_m$, showed a similar increase in the presence of 100 mM KCl. Inclusion of Ficoll 70, produced a further stabilization of apoCaM's structure (see Table 2), although the differences between bulk and maximum impact, at 400 mg/ml, were reduced because of the significant impact of ionic strength already noted.

## Results from computer simulations

The trend observed in the far-UV CD spectra in the presence of crowders is likely caused by structural changes in apoCaM. It motivated the investigation of CaM's structures using computer simulations and physical models that may explain the phenomena in responses to the changes in solutions such as ionic strength and the volume fraction of crowders ($\phi_c$). We focused on the structural analysis in the helical content, the EF-hand orientation, and the probability and correlation of contact formation among amino acids, particularly aromatic ones such as phenylalanine and tyrosine, in both apoCaM and holoCaM.

### *ApoCaM's folding energy landscape modulated by the ionic strength*

The radius of gyration ($R_g$) for apoCaM as a function of temperature in units of $k_BT/\varepsilon$ is provided in Figure S1A (see the supplement materials). At high temperatures, $R_g$ of apoCaM reduces as the ionic strength increases (See Figure S1). Interactions between partially charged amino acids in apoCaM were significantly screened at [KCl]=0.5M, allowing a compact unfolded ensemble structure. The profile of apoCaM's $R_g$ vs Temperature at [KCl]=0.5M nearly overlapped with our previous study using the coarse-grained protein model without the Debye-Hückel term. At a high temperature, 1.6 $k_BT/\varepsilon$, apoCaM's $R_g$ at [KCl]=0.1M can be 50% greater than the $R_g$ of apoCaM at [KCl]=0.5M.

The significant changes in apoCaM's radius of gyration by increased ionic strength prompted us to further characterize their structures using the Energy Landscape Theory [41,42]. At [KCl]=0.5M, a two-dimensional folding free energy landscape was constructed for apoCaM as a function of the overlap function [43] ($\chi$) and asphericity [44] ($\Delta$) at three different temperatures (Figure 4). $\chi$ measures the similarity to the structure obtained from the protein data bank (pdb) and it ranges from 0 to 1 where 0 is similar to the structure from the PDB. When similarity decreases, $\chi$ approaches to 1. $\Delta$ ranges from 0 to 1 where 0 represents a sphere and 1 represents a rod. The free energy basins represent the most probable structures of apoCaM. At T=1.07 $k_BT/\varepsilon$, apoCaM is populated in one basin named M1 in Figure 5A, which is defined in the ranges between 0.1<$\chi$<0.3 and 0.28<$\Delta$<0.3. The most probable structure in M1 represents an extended, dumbbell-like conformation and it is similar to the PDB structure of apoCaM (PDBID: 1CFD).

As temperature is further elevated to 1.15 $k_BT/\varepsilon$ (Figure 4B), there are two major basins. In addition to M1, another basin named M2 is populated. M2 is defined in the ranges between $0.3<\chi<0.4$ and $0.03<\Delta<0.15$. The most probable structure in M2 represents a compact, spherical-like conformation. When temperature is further increased to 1.4 $k_BT/\varepsilon$ (Figure 4C), only M2 exists.

The population shift between M1 and M2 was also shown to be dependent on the ionic strength. At T=1.15 $k_BT/\varepsilon$, Figure 5, the folding energy landscape shifts from populations of M2 (Figure 5A) to M1 (Figure 5C) when [KCl] increases, representing that attractions between charged amino acids are screened permitting an extended, dumbbell-like structure to become the most probable state. The screening effects also dominate the ensemble structures of the unfolded states. At a high temperature T=1.5 $k_BT/\varepsilon$, the unfolded state under a low ionic strength condition is found at the range of $0.7<\chi<0.8$ and $0.5<\Delta<0.7$ (Figure S2A). However, this basin vanishes at medium to high ionic strength. Instead, a structurally more compact and spherical ensemble emerges (Figure S2B, S2C), indicating that at high ionic strength, electrostatic interactions are screened and the unfolded state of apoCaM is thus destabilized.

*ApoCaM's ensemble structures depend on the ionic strength*

In order to investigate the structural detail of the M1 and M2 states, we analyzed their characteristics in terms of helicity, the orientation of EF hands, as well as the probability of contact formation between aromatic amino acids. At low [KCl] when M2 prevails, its helicity is 46.0%. At high [KCl] when M1 dominates, its helicity is 48.8%. This result suggests that when the ionic strength increases the population shifts toward the M1 state, and the helicity of apoCaM increases. The helicity of apoCaM at different ionic strengths as a function of temperature is shown in Figure 6A. Over a wide range of temperatures, the helicity of apoCaM at [KCl]=0.5M is up to 3% greater than that at [KCl]=0.1M and up to 2% greater than that at [KCl]=0.2M. These simulation results are consistent with the small but systematic increase in the negative bands in the CD spectra at 222 nm (predictive of α-helical content) as KCl increases (Figure 3). For a more direct comparison with the experimental data, We have computed a CD spectrum by including the calculated average helical and β−strand content from the simulations using a linear combination of the reference spectra [45] in Figure 6D. When the ionic strength increases, the negative bands in the calculated CD spectrum increase, which is qualitatively agreeable with the experiment data (Figure 3). Note that such calculations are based on oversimplifications of factors contributing to the CD spectra. For example, there is no available methodology to calculate contributions to the CD signal from tertiary structural components or rearrangements (for example, altered orientation of helices relative to each other).

We further evaluated the angles of the two helices (modeled as vectors) in each of the EF hands. The vector of an α-helix is defined by pointing a direction from the averaged position of the first four residues (N) to the averaged position of the last four residues (C). This follows a similar definition used previously [4]. The angle ($\Theta_{ij}$) between the two vectors drawn from the two helices, *i* and *j*, forming an EF hand is defined by the arccosine of the inner product of these two vectors. $\Theta_{ij}$ of apoCaM is different in each dominant ensemble structure in Table 3. Regarding the M1 state where the most probable structure adopts an extended conformation, $\Theta_{AB}$, $\Theta_{CD}$, $\Theta_{EF}$, $\Theta_{GH}$, and $\Theta_{EF}$ of each EF hand is individually much greater than a right angle, indicating that the alignment of the two helices in the EF-hands is nearly antiparallel. Regarding the M2 state where the most probable conformation resembles a collapsed compact structure, $\Theta$ of the same list of EF-hands become closer to a right angle. At a high ionic strength ([KCl]=0.5M), the population of M1 dominates, indicating that the ensemble average of $\Theta$ in each EF hands is closer to 180 degrees and the alignment of the two helices is nearly antiparallel.

In addition to $\Theta$, we also monitored the probability of contact formation among aromatic residues (phenylalanine and tyrosine) of apoCaM in M1 and M2 states (Figure S3A and S3B, respectively). We found that the probability of such contacts was affected by the reorientation of helices of the EF hands in response to different ionic strengths. The probability of contact formation between aromatic residues is plotted as a matrix and the color is scaled from 0 to 1. The matrix elements at the upper triangle denote the contacts of a structure modeled from apoCaM's PDB structure (PDBID: 1CFD), while those in the lower triangle denote the contacts found in the states other than the PDB one (nonnative contacts). There are two groups of contact formation that can reflect the structural differences between M1 (extended, Figure S3A) and M2 (collapsed, Figure S3B) of apoCaM. The first group consists of contacts formed by Phe12 and Phe16 at Helix A and Phe65 and Phe68 at Helix D in the N-terminus (circled in green). The second group consists of contacts formed by Phe89 at Helix E and Tyr138 and Phe141 at Helix H in the C-terminus (circled in orange). There is a high probability of contact formation of these two regions in the M1 state, indicating the contacts between Helix A and D in the N-lobe and the contacts between Helix E and G in the C-lobe are highly similar to the structure modeled from apoCaM's PDB structure (PDBID: 1CFD). However, the probability of maintaining these contacts decreased by 5-fold in the M2 state, allowing the aromatic residues to break free from each other forming non-native contacts between the N- and C-domains. Thus, formation of nonnative contacts are prevalent in the M2 state, demonstrated by a noticeable growth of contact formation in the lower triangle in Figure S3B.

***ApoCaM's folding energy landscape is modulated by crowding agents***

The radius of gyration ($R_g$) of apoCaM in the presence of crowders as a function of temperature in units of $k_BT/\varepsilon$ is provided in Figure S1A. Across the spectrum of ionic strengths examined, at high temperatures, $R_g$ was reduced at high $\phi_c$. A more compact form of unfolded states produced by macromolecular crowding agrees with former studies [24,27,46,47] in which the stability of the folded state of a protein can be relatively enhanced [13,48,49,50]. At [KCl]=0.5M, at T=1.07 $k_BT/\varepsilon$, the dominant structure of apoCaM in the presence of crowders is M1 (Figure 4D), while at T= 1.4 $k_BT/\varepsilon$, the dominant structure is M2 (Figure 4F). At the onset of transition (T=1.15 $k_BT/\varepsilon$), the population of M2 increases in the presence of crowder (Figure 4E), in comparison to that in the absence of crowders (see Figure 4B). In addition, the population shift between M1 and M2, at T=1.15 $k_BT/\varepsilon$, was also shown to be dependent on the ionic strength in the presence of high $\varphi_c$ (Figure 5). We plotted the same folding energy landscape as Figure 4B and 4E but varying the ionic strengths for [KCl]=0.1M (Figure 5A and 5B) and [KCl]=0.2M (Figure 5D and 5E). Overall, M2 is statistically more probable in the presence of crowders across the spectrum of ionic strength because M2 is more spherical and compact over M1.

*Helicity of apoCaM in the presence of crowding agents*

The helicity of apoCaM in the presence of crowders at different ionic strengths as a function of temperature is shown in Figure 6B. Over a wide range of temperatures, the helicity of apoCaM at [KCl]=0.5M was up to 3% greater than that at [KCl]=0.1M and up to 2% greater than that at [KCl]=0.2M. Interestingly, for each ionic strength examined (0.1, 0.2, 0.5M), there was little difference between the helicity of apoCaM in the absence and in the presence of crowders (Figure 6A&B). Correspondingly, we computed a CD spectrum using the content of helices and β-strands in the simulations in the presence of crowders in Figure 6D. Surprisingly, the two computed CD spectra for bulk and $\varphi_c$=40% overlay on top of each other. This finding does not agree with experiments in Figure 3, indicating that there exist other reasons other than the helicity that could impact CD spectra when crowders are present. As noted above, the calculated CD spectra account only for α-helical and β-sheet content and ignore important structural rearrangements such as reorientation of helices that contribute to the experimentally measured CD spectra. We propose that this is the explanation for why the calculated CD spectra are different than experiments under crowded conditions and warn against over-interpreting calculated CD spectra that do not account for all sources of altered CD signals measured experimentally.

*HoloCaM's folding energy landscape is not affected by the ionic strength*

The radius of gyration ($R_g$) for holoCaM as a function of temperature in units of $k_BT/\varepsilon$ was provided in Figure S1B (see the supplement materials). Overall, the structures remain compact at high and low ionic strengths. A two-dimensional free energy landscape as a function of overlap function ($\chi$) and asphericity ($\Delta$) of holoCaM in different temperatures was represented in Figure 7. The most probable state, M3, of holoCaM is defined between $0.38<\chi<0.45$ and $0.01<\Delta<0.1$. In addition, this is also the most probable state existing in a wide temperature range (Figure 7) as well as in different ionic strength conditions (Figure S5). Thus, the dominant state of holoCaM is M3 and this state is very stable.

*Characterization of holoCaM's ensemble structures*

In order to investigate holoCaM's structure in the M3 state, we analyzed its characteristics in terms of helicity, the orientation of EF hands, as well as the probability of contact formation between aromatic amino acids as we did for apoCaM. The helicity of M3 is 49.7%. Figure 6C shows the helicity of holoCaM at different ionic strength as a function of temperatures and there is little difference. Interestingly, holoCaM has a greater helicity than apoCaM by 5% at high temperature.

The EF-hand orientation of holoCaM's most probable structures in M3 is shown in Table 3. $\Theta_{AB}$ between Helix A and B in the structures of M3 are at a right angle, indicating that their alignment is nearly perpendicular to each other. Similarly, the orientation between Helix C and D, Helix E and F, Helix G and H, is perpendicular, unlike those of M1.

The probability of contact formation, involving the aromatic residues phenylalanine and tyrosine, of holoCaM in the M3 state is represented in Figure S4. The probability of contact formations involving Phe12 and Phe16 at Helix A and Phe65 and Phe68 at Helix D in the N-terminus (circled in green) and the contacts formed by Phe89 at Helix E and Tyr138 and Phe141 at Helix H in the C-terminus (circled in orange) is around 0.5. Contacts across domains are evident. HoloCaM is structurally robust and compact, compared to apoCaM.

*Covariance analysis of contact formation of apoCaM and holoCaM*

Experiments had demonstrated that the dynamics of the N- and C-lobes impact one another and they are not isolated [38,40,51,52]. It was speculated that electrostatic interactions contributed, in part, to the cross-domain dynamics[38]. A mutation at the C-lobe that decreased its affinity to calcium negatively impacted the calcium binding ability of the N-lobe [52]. Inspired by this, we computed a covariance matrix to investigate the correlation of contact formation between the N- and C-lobes at T=1.15 $k_BT/\epsilon$ under various ionic strengths for holoCaM and apoCaM. We found that the interdomain contacts in the N- and C-domains of apoCaM are correlated (boxed in orange in Figure 8). At [KCl]=0.1M these contacts become more negatively correlated (Figure 12A), while at [KCl]=0.5M, these same contacts become positively correlated (Figure 12B). However, for holoCaM, these contacts are always negatively correlated in both low and high ionic strength conditions (Figure S6).

We analyzed the structural proximity of these correlated contacts in terms of helices as shown in a schematic illustration in Figure 9. We strategically picked four representative contact pairs. *a* represents the contact (Phe12 and Met76) between helix A and helix D, *b* represents the contact (Ala1 and Ala147) between helix A and helix H, *c* represents the contact (Phe89 and Phe141) between helix E and helix H, and *d* represents the contact (Ala1 and Glu83) between helix A and helix E. *a, b*, and *c* are the native contact pairs found in the apoCaM PDB structure, *d* is a non-native contact pair that is made in the collapsed form of apoCaM at low ionic strength.

At [KCl]=0.5M (most populated around the M1 ensemble), contacts *a* (located in the N-lobe) and *c* (located in the C-lobe) are positively correlated ($cov_{ac}$=0.25). The interdomain correlation between *a* and *c* is established through an interdomain contact *b*, where $cov_{ab}$ =0.32 and $cov_{bc}$ =0.39. If b is formed across the two domains, then this contact formation positively contributes to the formation of contact *a* and *c*. Thus, contacts a and c are positively correlated. This is illustrated in Figure 9A.

At [KCl]=0.1M (most populated around the M2 ensemble), contacts *a* and *c* are now negatively correlated ($cov_{ac}$= -0.27). As a result, a nonnative contact pair *d* forms between helix A and helix E. Subsequently $cov_{ad}$ = 0.60, the formation of the nonnative contact pair *d* is strongly tied to the formation of native contact *a*. As a result, a core structure is formed by helix A, helix D and helix E, and it excludes the interdomain contact *b*. In our simulation the probability of contact formation of b is close to zero. Without *b* in place Helix H becomes flexible and it avoids the contact formation to Helix E, which suppresses the formation of contact *c*. Thus, contacts *a* and *c*, despite being intradomain contacts, are negatively correlated. This is illustrated in Figure 14B. Since the dominant state of holoCaM is M3 (a compacted structure) in both low and high ionic strength conditions, the contacts defined above are always negatively correlated (Figure S6).

We also investigated the contact correlation associated with Asp93 and Asp129, studied by the previous NMR work [52] about the presence of correlated interactions between N-lobe and C-lobe. We studied these contacts of apoCaM at high ionic strength where charge-charge interactions are screened and interactions between domains become less dependent on a specific charge on amino acids. There are a greater number of contacts in Asp93 than Asp129 such that the correlations of contacts pairing with Asp93 and contacts in the N-lobe are stronger, so we focused on Asp93. When we traced the contact formation of the Contact Pair Index (CPI) 204 formed by Asp93 and Ile100 (Table S3) in Figure 12B, CPI 204 is located inside the orange box where there is a positive correlation with the contacts in the hydrophobic pocket of the N-lobe (e.g., CPI 28 involving Phe12 and Phe65). In contrast, in the simulations on holoCaM, there is negligible correlation between contacts associated with Asp 93 and Asp 129. For example, CPI 175 (involving Asp93 and Ile100; see Table S4) and contacts in the hydrophobic pocket of N-lobe (CPI 18, involving Phe12 and Phe 65; see Table S4) are weakly correlated, implying that structural fluctuations in C-domain impacts less on the exposure of hydrophobic pocket in N-domain.

## Discussion

*Conformations of apoCaM and holoCaM in aqueous solution*

Our simulation results indicate that apoCaM has both extended, dumbbell-like (M1) and compact, spherical-like (M2) structures in aqueous solution, consistent with published experimental results [53]. When the alignment of EF-hands is antiparallel, this orientation represents a "closed" state for target binding and the hydrophobic residues that dominate the initial interactions with protein targets are buried inside the lobes of apoCaM. Subsequently, the probability of contact formation among selected hydrophobic residues (e.g. aromatic amino acids) exceeds 90% (Figure S3).

We compared our analysis from simulations of apoCaM in the M1 state to experimental findings based on NMR measurements [4]; see Table 3. $\Theta_{AB}$, $\Theta_{CD}$, and $\Theta_{GH}$ are close to 180 degrees which are in close agreement with experiments. Although $\Theta_{EF}$ is off from experimental measurement, this may be due to the instability of the C-lobe relative to the N-lobe. This particular EF-hand whose $\Theta_{EF}$ is close to a right angle may represent a semi-open state of the C-lobe which would be consistent with a NMR study examining the structure of the C-lobe of CaM [54].

Regarding an ensemble of holoCaM structures, there is a remarkable lack of heterogeneity in conformations with only one compact state (M3) dominating in solution. In a coarse-grained holoCaM protein model, each of the four calcium beads is "fixed" into an EF-hand motif through springs. As a result, unlike apoCaM, the basin of free energy landscape of a holoCaM is quite narrow. In the M3 state, the alignment of two helices in each EF-hand is perpendicular to each other (Table 3). This perpendicular orientation of the helices represents an "open" state for target binding and the hydrophobic residues are readily exposed. A reorientation of these helices has been suggested to lead to increased negative bands in the CD spectrum [38] and is consistent with our data comparing the Far-UV CD signal of holoCaM and apoCaM (Figure 2).

Note that M1 and M2 are the two dominant ensemble structures from the simulations of apoCaM. M3 is the most dominant ensemble from the simulation of the holoCaM. There is no direct correlation of these ensemble structures to the three-state model (N, U, I) for fitting the folding stability of apoCaM in experiments. The latter captures the known independent unfolding of the two lobes of CaM during the thermal denaturation process[40].

*Impact of ionic strength on apoCaM and holoCaM*

We investigated the effect of ionic strength on the structural changes in apoCaM and holoCaM in which 30% of the amino acids in CaM are highly charged residues such as Arg, Lys, Glu, and Asp. Regarding holoCaM, the presence of four calcium beads constrains the loops of the EF-hands to such an extent that further structural changes in response to ionic strength increases is trivial (Figure S5). This is consistent with the dominant effect of calcium binding over ionic strength on the resistance of holoCaM to thermal denaturation (compare Figures 2).

However, apoCaM is a malleable protein and its conformation can be easily manipulated by perturbations in the environment [17]. Electrostatic interactions between the charged residues, if not screened, produce an instability of apoCaM's native state that resembles an extended dumbbell shape (M1). However, destabilizing electrostatic interactions between like-charges can be sufficiently screened out by increased ionic strength such that the population of the extended M1 state becomes highly probable; hence, the helicity of apoCaM increases (Figure 6A). This may explains why the negative band of far-UV CD signal of apoCaM increases with the ionic strength (Figure 3).

*Effect of macromolecular crowding on apoCaM*

The investigation of macromolecular crowding effect on apoCaM was initiated in our previous work [17]. Under crowded conditions that exert excluded volume interactions on apoCaM, compact forms are statistically most probable. As a result, the stability of the compact state of apoCaM increases with crowding level (Fig S1A), which is in agreement with experiments (Fig 3). Interestingly, in our protein model of apoCaM that allows attractions between nonnative interactions, there could exist multiple compact states with various shapes. As a result, at high volume fraction ($\phi_c$) of crowding agents, the shape of compact proteins strongly influences the population shifts among themselves [12,13,55]. For example, the M2 state is structurally more spherical than M1. In the presence of a high volume fraction of crowders, the population of the M2 state (spherical) is more probable than the M1 state (dumbbell-like) because the former presents a smaller covolume as a sphere, as indicated by the Scaled Particle Theory (SPT) [55]. This conclusion is also well supported by computer simulations on other anisotropic proteins under high levels of macromolecular crowding [12,13]. As a result of population shifts towards a collapsed spherical M2 state under macromolecular crowding effects, while there is little impact on the helicity, the alignment of two helices in an EF-hand adopts nearly right angle conformations as shown in the M1 state (Figure 6AB), which may contribute to an increase of the far-UV CD signal (Figure 3). Although both ionic strength and crowding levels cause an increase in negative bands in the far-UV CD signal, we propose that the unique physical-chemical differences in these two conditions can favor different compact structures of apoCaM that give a similar kind of phenomenological response in the CD experiments.

*Effect of macromolecular crowding and ionic strength on apoCaM*

Our findings on CaM suggested that the ensemble of this highly charged protein is strongly dependent on the ionic strength and the macromolecular crowding effect. In CaM, the number of negatively charged residues (38) is greater than that of positively charged residues (14). A strong repulsion produced results in a higher probability of finding apoCaM in different states. At higher ionic strength such electrostatic repulsions are screened and the stability of CaM increases. This trend has also been found experimentally in other proteins[56]. At high levels of macromolecular crowding, the compact structures are populated due to depletion-induced attraction as a result of particle fluctuations, an entropic effect explained by Asakuwa and Oosawa[57,58]. A competition between electrostatic interactions and macromolecular crowding effects will depend on the chemical nature of amino acid sequences and the structure of CaM. Due to the charge distribution of CaM, the specific details of our findings may be unique to this protein, however, evaluating both electrostatics and crowding as done in the present study is clearly needed to fully evaluate the impact of these factors on protein structures. Interestingly, our study may be applicable to the investigation of other biopolymers that are structurally malleable and highly charged, such as RNA[59] and intrinsically disordered proteins[60].

*Interdomain interaction between N- and C-lobes*

Experiments demonstrated that the N- and C-lobes of CaM impact one another and the two lobes cannot be treated as isolated entities [38,40,51,52]. It was speculated that electrostatic interactions contributed to the cross-domain dynamics [38]. Interestingly, in a recent findings with NMR [52], mutations from Asp to Ala of CaM at a $Ca^{2+}$-binding site in the C-lobe affect the target affinity of the N-lobe. However, there has not been a systematic analysis of contacts that might provide explanations for these results.

Our *in silico* results indicate that there exist strong interdomain dynamics demonstrated by correlated contact formation in the N- and C-lobes (Figure 8). Interesting, N- and C-lobes are positively correlated at 0.5M ionic strength and negatively correlated at 0.1M. Notably, although the ionic strength is known to screen out electrostatic interactions, the contact formations that are affected by ionic strength in our simulations are hydrophobic residues (Phe12, Met76, Phe89, Phe141; see Figure 9). Thus, the electrostatic interactions appear to indirectly, rather than directly, affect the correlation between the lobes. We propose that the conformation of CaM, induced by different ionic strength, is the key to control the correlation between two lobes. In high ionic strength when interactions from like charges are sufficiently screened out, CaM is a dumbbell-like structure; N- and C- lobes are positively correlated contributed by the interdomain contacts between Helix A and Helix H (yellow contacts in Figure 9A). However, in low ionic strength, CaM is a compact structure. As a result, a core structure is formed by helix A, helix D and helix E, and it excludes the interdomain contact between Helix A and Helix H (yellow contacts in Figure 9B). Accordingly, the contacts between Helix E and Helix H breaks (green contacts in Figure 9B). The overall result is the anticorrelation between N- and C-lobes.

Our simulations of apoCaM at high ionic strength supports the conclusions of previous NMR work [52] about the presence of correlated interactions between N-lobe and C-lobe through inter-domain interactions. In these experimental findings [52], Asp to Ala mutations at the $Ca^{2+}$-binding sites in the C-lobe (Asp93 and Asp129) altered the conformation of the C-lobe as well as the hydrophobic pocket of the N-lobe. This results in an increase of the $Ca^{2+}$-binding affinity of the N-lobe by decreasing the dissociation rate of $Ca^{2+}$ ($k_{off}$). Our findings suggested that there is a strong positive correlation between contacts associated with Asp93 (e.g., CPI 204) in the C-lobe and the hydrophobic pocket in the N-lobe (e.g., CPI 28). We can speculate that if there is a mutation to replace Asp93 to Ala that weakened CPI 204, CPI 28 involving Phe12 and Phe65 is similarly weakened. When contacts in the hydrophobic pockets are weakened, the alignment between the two helices in an EF-hand will resemble holoCaM, instead of apoCaM, favoring the retention of calcium ions. In addition, a holoCaM-like conformation indicates an increase in the solvent exposure of hydrophobic amino acids, which is in consistent with an increase of ANS-binding of CaM mutants [52].

*Charge-charge interactions in coarse-grained protein model*

Our combined experimental and computational effort highlights the importance of ionic strength and crowding effects on apoCaM's conformations. In the computational portion, we developed a novel algorithm to assign charges on the coarse-grained protein model in response to the absence and presence of calcium ions by combining the methods of coarse-grained molecular simulations, protein reconstruction, quantum chemistry, and statistical physics. We have discussed the charges computed by MOPAC with another high-level quantum mechanics/molecular mechanics (QM/MM) at a Hartree-Fock level in the Supplement. The two sets of charges focusing on a part of CaM are in very good agreement and it justified the use of semi empirical calculation for this study, given that it is beyond the capability of current computing resources to treat the entire CaM as the QM region for this study. To the best of our knowledge, there is no experimental data to compare with the calculations in our model. In addition, computer simulations including calcium have been limited to a small portion of CaM [18] (e.g. a loop in an EF-hand) and unable to address the dynamics of an entire protein. Our method has provided a reasonable starting point to address the charge distribution on CaM in the absence and presence of calcium averaged over an ensemble when the conformational fluctuations were taken into account.

In contrast to our previous coarse-grain protein model [61] a Debye-Hückel potential is included in the Hamiltonian in order to investigate the electrostatic interactions under different ionic strength conditions. Debye-Hückel potential has been successfully employed in many systems where interactions between charged particles are critical to biological functions, such as protein sliding along DNA [62], RNA folding [32], structures of intrinsically disordered proteins [60], and dynamics of chromatin fiber formation [63]. However, assignment of charge distribution in these systems has been approached in a qualitative fashion. For example, an arbitrary +1 or -1 can be assigned to charged amino acids such as Arg, Lys, Glu, and Asp in proteins [62]. In other cases, charges on coarse-grained model were derived from all-atomistic force fields [64]; however, charges from all-atomistic force fields have yet to include the effect of bound ion in proteins [18]. It could be problematic when specific charge-charge interactions, such as the binding of calcium ions, are critical in CaM's dynamics and target recognition. In this report, we describe a new method that evaluates a reasonable charge distribution of a coarse-grained protein in order to investigate the relationship of conformational changes and the binding of ions in a protein (such as CaM).

The challenge in this problem arises from the fact that the charges on atoms are dependent on the extent of their solvent exposure in an ensemble of structures. Our strategy was to compute the charge distributions of CaM in solutions by averaging over an ensemble of apoCaM or holoCaM conformations. To our best knowledge, we are the first to deploy a combined method of coarse-grained molecular simulation that produces an ensemble of structures, protein reconstruction and quantum chemistry methods that generate charges in atomistic details. Statistical physics is then employed to evaluate a reasonable average of charges on coarse-grained CaM proteins. With this combined approach, we now have a powerful tool to evaluate the conformations of apoCaM or holoCaM in response to changes in the physical-chemical nature of the environment. We are aware of potential pitfalls of this model, which we will address in future studies.

These issues include the excessive consideration of electrostatic interactions reflected in both the Debye-Hückel potential and the London dispersion force (due to dipole-dipole interaction of charged particles) in the Lennard-Jones potential. Despite these concerns, we believe our new method provides the most accurate simulation based approach to address the effects of ionic strength on protein structure. At high ionic strength (e.g. [KCl] is 0.5M), because the electrostatic interactions are sufficiently screened, the energy sum contributed from the Debye-Hückel potential is sufficiently small and negligible when compared to the sum from the Lennard-Jones potential. Indeed, under this condition, the structural analysis of apoCaM based on a Hamiltonian including both the Debye-Hückel potential and the Lennard-Jones potential converges with the structural analysis of apoCaM based on a Hamiltonian including only the Lennard-Jones potential. At low ionic strength, because the charge-charge interactions are less screened, the energy contribution the Debye-Hückel potential is only 9% of the Lennard-Jones potential. Such an increase of electrostatic interactions qualitatively reflects the correct physics between charged particles in solutions when the concentration of ions is low.

## Methods

*Experiments:*

*Expression and Purification of CaM*

The expression and purification of CaM was exactly as described previously[65]. Briefly, CaM was produced from the expression plasmid pCR2 that encodes a full-length cDNA of a codon-optimized version of CaM that is identical in amino acid sequence to vertebrate CaM. CaM was produced from the plasmid in the BL21(DE3)star *e. coli* strain (Invitrogen) using standard IPTG induction. After 4-6 hr of induction, the cells were harvested, lysed in hypotonic buffer, and the CaM was purified through a series of ammonium sulfate precipitation and phenyl-Sepharose chromatography steps as described[65]. The CaM from the last chromatography step was dialyzed into 50 mM MOPS, pH 7.0 and stored as aliquots at -80ºC. Analysis of these preps by SDS-PAGE showed the protein was greater than 95% pure and the protein concentration was determined by $A_{280}$ nm (using a molar extinction coefficient of 2980; 1 $A_{280}$ = 5.65 mg/ml) and BCA assay (Pierce), confirmed by quantitative amino acid analysis.

*Circular Dichroism*

Far-UV CD spectra were collected on a JASCO-815 spectrophotometer controlled by Spectra Manager software. Suprasil cuvettes with a 1.0 mm path length were used for all experiments. The spectrometer parameters were typically set to the following unless noted otherwise: band width, 1 nm; response time, 4 s; data pitch, 0.2 nm; scan speed 50 nm/min and two acquisitions were averaged for each measurement. Addition of Ficoll 70 (as crowder) contributed significantly to the background absorbance at wavelengths below 205 nm and this area of the scans was found to not be reliable for data fitting. Buffer only background scans were collected and subtracted from the experimental data for each condition. To produce thermal denaturation curves, temperature rate changes of 0.5°C/min, 1°C/min and 2°C/min were assessed at 220 nm. No significant differences were noted between these temperature rate changes and all data presented in this manuscript was collected at 1°C/min.

The base buffer for these experiments was 10 mM MES, pH 7.0. To produce data for apo-CaM, EDTA was added to a final concentration of 1 mM while maintaining a pH of 7.0. For holo-CaM ($Ca^{2+}$-saturated), EDTA was omitted and $CaCl_2$ was added to a final concentration of 1 mM. Ficoll-70 PM (Sigma/Aldrich) was used as a crowding agent and was varied between 100-400 mg/ml. To produce these solutions, a 400 mg/ml solution in base buffer was first prepared and residual insoluble material was removed by filtering through sequential 5 μm and 0.2 μm syringe filters. Then decreasing concentrations of Ficoll were prepared by diluting the 400 mg/ml stock solution with base buffer to the desired final concentrations. In some experiments, the impact of ionic strength was assessed by inclusion of KCl in a range between 0 and 500 mM final concentration. In initial experiments, CaM concentration was varied between 0.1 and 0.8 mg/ml. A concentration of 0.4 mg/ml (~25 μM) was found to provide excellent signal/noise in the CD spectra and was chosen for the experiments presented in this manuscript. Each experimental condition was assayed in duplicate, and the results averaged.

*Analysis of CD data*

The CD signal in the raw data is recorded in millidegrees. For normalization and data fitting, the raw data was converted to mean residue ellipticity by the following formula, Ellipticity, [θ], in deg*cm$^2$/dmole = (millidegrees * molecular weight)/(pathlength in millimeters * concentration of protein in mg ml$^{-1}$ * the number of residues). The molecular weight of CaM is 16706 Da, the path length is 1 mm, the protein concentration is 0.4 mg ml$^{-1}$, and the number of residues of CaM is 148.

The thermal denaturation data were fit to two models of unfolding in Origin software. First, a two-state model describing unfolding as a single transition between a native (N) and an unfolded (U) state was employed using the following formula:

$$Y = (I_N + S_N T)f_N + (I_U + S_U T)f_U$$

To distinguish changes in conformation of the N- and C-lobes of CaM, a three-state model describing the unfolding of protein in two transitions: native (N) denatures to an intermediate (I) which then denatures to an unfolded (U) state was employed and is described by the following formula:

$$Y = (I_N + S_N T)f_N + (I_I + S_I T)f_I + (I_U + S_U T)f_U$$

where $T$ is the temperature. $I_i$ and $S_i$ are intercepts and slopes for the optical signals of native, intermediate, and unfold states by assuming linear behavior at these states. $f_i$ are the fractions of the total population in each state. The fractional populations were determined from the equilibrium constant for unfolding, $K_u$ (= $\exp(-\Delta G_T/RT)$). Free energy values used in calculating the fractional populations were obtained from the modified Gibbs-Helmholtz equation:

$$\Delta G_T = \Delta H_m \left(1 - \frac{T}{T_m}\right) - \Delta C_p(T_m - T + T \ln\left(\frac{T}{T_m}\right))$$

where $T_m$ and $\Delta H_m$ are the temperature of the transition midpoint (K) and the van't Hoff enthalpy at $T_m$, respectively, and $\Delta C_p$ is the difference in heat capacity between native and unfolded states. In fitting the unfolding data for holoCaM to a two-state model, all parameters were allowed to vary. To obtain reasonable fits of the data of apoCaM to a three-state model, $\Delta C_p$ had to be fixed at 0.

*Computer Simulation:*

*Coarse-grained modeling of apoCaM and holoCaM proteins*

A side chain $C_\alpha$ model (SCM)[61] that includes two beads per amino acid (except glycine) was used to represent the protein structure of CaM. A detailed description of the Hamiltonian of this model for apoCaM (without assigned partial charges) can be found in our previous work [17]. Nonbonded interactions between side chain beads follow the Betancourt-Thirumalai statistical potential[66].

The protein part of the calcium bound CaM followed the same aforementioned coarse-graining procedure as apoCaM. The interactions between calcium ions and partially charged amino acids will be discussed in the next section. The $Ca^{2+}$ ion in each EF-hand of holoCaM (PDBID: 3GOF) was represented by a bead: $CA_m$ (m=1,2,3,4). Each $CA_m$ was attached to the coarse-grained side chain beads of the nearest negatively charged amino acids through spring-like interactions (Eqn 1). $CA_1$ was connected to Asp20, Asp22, and Asp24, Glu31; $CA_2$ was connected to Asp56, Asp58, Asp64 and Glu67; $CA_3$ was connected to Asp93, Asp95, and Glu104; $CA_4$ was bonded to Asp129, Asp131, Asp133 and Glu140. The bond energy, $E_{bond}$, is represented as:

$$E_{bond}^{mj} = k_b(r^{mj} - r_0^{mj})^2 \quad (m=1,2,3,4) \qquad \text{Eqn. (1)}$$

where $m$ represents $CA_m$ and $j$ is the coarse-grained side chain bead of the amino acids connected to $CA_m$. $k_b = 100\varepsilon$ and $\varepsilon = 0.6$ kcal/mol. $r_0^{ij}$ was calculated from the crystal structure of holoCaM (PDBID: 3GOF).

The bond-angle energy term, $E_{angle}$, involving $CA_m$ consisted of three consecutive beads $i$, $m$, and $j$, where $m$ represents $CA_m$ and $i, j$ are the coarse-grained side chain beads of the residues bonding to $CA_m$. $E_{angle}$, is in the following form:

$$E_{angle}^{imj} = k_\theta(\theta^{imj} - \theta_0^{imj})^2 \quad (m=1,2,3,4) \qquad \text{Eqn. (2)}$$

where $k_\theta = 20\varepsilon$ and $\theta_0^{imj}$ was calculated from the crystal structure of holoCaM.

The nonbonded interactions, $E^{im}$, between any bead of the coarse-grained protein, $i$, and $CA_m$, are strictly repulsive to represent only the steric effect (Eqn. 3), where $\varepsilon = 0.6$ kcal/mol and $\sigma_{ij} = f \cdot (\sigma_i + \sigma_j)$; $\sigma_i$ and $\sigma_j$ are the Van der Waals radii of the interacting beads. $f=0.9$ is a scaling factor to avoid steric clashes between interacting beads.

$$E_{steric}^{im} = \varepsilon \left(\frac{\sigma_{ij}}{r_{ij}}\right)^{12} \qquad \text{Eqn. (3)}$$

*Inserting missing residues of holoCaM*

There are a total of four amino acids missing in the PDB file of holoCaM (3GOF): the first two amino acids in the N-terminus (Ala1 and Asp2) and the last two amino acids in the C-terminus (Ala147 and Lys148). We inserted these four missing amino acids to holoCaM by using an in-house reconstruction method, side-chain $C_\alpha$ to all-atom (SCAAL) method[12,37]. The structure of apoCaM (1CFD) was employed as a template and the coarse-grained SCM structure of holoCaM (3GOF) was used to restrain the positions of $C_\alpha$ and side-chain beads. A part of the structural Hamiltonian of the missing residues, such as the bond and bond-angle energy terms, adopted the same formula and the same parameters from those of apoCaM. However, to reserve its polymeric flexibility at the termini, dihedral energies of these missing residues were not considered by setting related parameters to 0.

*Assigning charge distributions by combining MultiSCAAL and Quantum Chemistry*

The charge distribution in CaM was strongly affected by the presence of calcium ions, the charged states of their N- and C- termini, and the extent of solvent exposure in a structure. To determine the charge distribution of coarse-grained CaM, we executed the following steps, involving the method of statistical physics and quantum chemistry, as follow: (1) Produced an ensemble of coarse-grained apoCaM (or holoCaM) protein structures without charged amino acids at 1.13 $k_BT/\varepsilon$ from a conventional coarse-grained molecular simulation[26]; (2) 400 uncorrelated coarse-grained structures were selected from such ensembles using a Metropolis criteria[67] which depend on the potential energy difference between any randomly selected structure in an ensemble and the structure with the lowest potential energy in the free energy landscape. They were further reconstructed to all-atomistic protein representation using SCAAL [12,37]; (3) the termini of apoCaM (or holoCaM) were chemically capped by using a software VMD[68] and its built-in CHARMM force field [69,70] such that their charges were neutralized. The N-terminus was acetylated and the C-terminus was capped by -COOH which matched the condition of a modified calmodulin in experiments [1]; (4) the partial charge on every atom in a apoCaM(or holoCaM) structure was computed by using a semi-empirical quantum chemistry program MOPAC [71] based on AM1. The charge distribution of an ensemble of apoCaM (or holoCaM) structures was then computed by averaging over 400 selected structures giving negligible errors; (5) The average charge distribution of all-atomistic CaM was assigned to its coarse-grained representation including a C$\alpha$ bead that carries the total charge from the backbone atoms and a side-chain bead that carries the total charge from the side-chain atoms. The tabulated partial charges of coarse-grained apoCaM and holoCaM are provided in the Supplement as Table S1 and Table S2, respectively.

*Interactions between calcium ions and partially charged amino acids*

Debye-Hückel potential [32,36,62](Eqn. 4) was used to represent the interaction between two single partial charges in solutions with given ionic strength. The relative dielectric constant was chosen to be 80 for aqueous solution. The Debye-Hückel potential, $V_{ij}$, between beads *i* and *j* follows the following formula,

$$V_{ij} = \frac{Z_i Z_j}{4\pi r \varepsilon_o \varepsilon_r} e^{-r/\sqrt{\varepsilon_r \varepsilon_o k_B T / 2e^2 I}}$$

Eqn. (4)

$Z_i$ ($Z_j$) is the partial charge on each bead *i* (*j*). *r* is the separation between bead *i* and *j*. $\varepsilon_o$ is the permittivity of free space. $\varepsilon_r$ is the relative dielectric constant (set to be 80 representing the aqueous solution). $k_B$ is the Boltzmann constant. $T$ is the temperature. $e$ is the charge of an electron, and $I$ is the ionic strength of the aqueous solution.

*Simulation method*

Coarse-grained molecular simulations were performed on apoCaM and holoCaM, respectively under different macromolecular crowding levels. Ficoll 70, a semi-rigid crowding agent[72], was modeled as a hard sphere with a radius of 55Å that provides volume exclusion via nonspecific repulsions. The size of a cubic box in the simulation for the investigation of macromolecular crowding effects on CaM is 1140Å and the periodic boundary condition was used. The cutoff of any nonbonded interaction was set to half of the box size. The thermodynamics properties of apoCaM and holoCaM in bulk and in crowded environments (the volume fraction of Ficoll, $\phi_c$, is set to 25% and 40%) were investigated through the Langevin equations of motion at low friction limit [73]. The Replica Exchange Method [74,75] was implemented to enhance the sampling efficiency where simulations were performed concertedly at 18 temperatures ranging from 1 $k_BT/\varepsilon$ to 1.63 $k_BT/\varepsilon$. The exchange ratio was between 0.2 and 0.3. The integration time step is $10^{-4}\tau_L$, where $\tau_L = (m\sigma^2/\varepsilon)^{0.5}$ (m is the mass of particle, σ is the van der Waals' radius of the amino acid residues and ε is the solvent-mediated interaction energy). The data is sampled at every 1000 integration time step. Each exchange was attempted at every 400$\tau_L$. At least 40000 statistically uncorrelated data points were collected at each temperature. Thermodynamic properties and errors were calculated using the weighted histogram analysis method [76].

*Covariance matrix analysis*

The covariance matrix ($cov_{ij}$) represents the correlation among the formation of contact pairs *i* and *j*. The covariance is defined as following:

$$cov_{ij} = \frac{<q_i q_j> - <q_i><q_j>}{<q_i^2> - <q_i>^2}$$

where $q_i$ and $q_j$ is the probability of contact formation of the *ith and jth* contact, respectively. <> represents the ensemble average. $cov_{ij}$ ranges from -1 (negatively correlated) to 1 (positively correlated).

# Acknowledgements

MSC thanks the computational sources from the Texas Learning and Computation Center and Research Computing Center at the University of Houston.
22

**Figure legends**

Figure 1. Cartoon representations of calmodulin in the absence and presence of calcium ions. Cartoon representations are colored from the N-terminus (red) to the C-terminus (blue). (A) apoCaM. (B) holoCaM and the calcium ions are represented by yellow spheres.

Figure 2. Far-UV CD spectra and thermal stability. (A) Comparison of the far-UV CD spectra of apoCaM ($Ca^{2+}$ free) and holoCaM ($Ca^{2+}$-saturated). (B) Comparison of the thermal stability of apoCaM and holoCaM. The base buffer was the same as in panel (A) and the CD signal was monitored at 222 nm while the temperature was ramped from 5 to 90°C at a rate of 2.5°C/min. Note that under these conditions, apo-CaM exhibited a complex unfolding profile best fit with a 3-state model. In contrast, holoCaM did not fully denature even with temperatures up to 90°C. The unfolding profile of holoCaM was best fit with a 2-state model. The base buffer for these experiments was 10 mM MES, pH 7.0, that contained either 1 mM EDTA or 1 mM $CaCl_2$. The CaM concentration was 0.4 mg/ml

Figure 3. The impact of ionic strength and macromolecular crowding on the thermal stability of apoCaM. (A) Far-UV spectra were collected for apoCaM in base buffer, 10 mM MES, pH 7.0 with 1 mM EDTA, at increasing ionic strength by inclusion of KCl from 50 mM to 500 mM. (B) Thermal denaturation (CD at 222 nm) of apoCaM in the presence of increasing KCl from 50 to 500 mM. (C) Far-UV CD spectra of apoCaM in the presence of Ficoll 70 at crowding levels from zero to 400 mg/ml. (D) Thermal denaturation (CD at 222 nm) of apoCaM in the presence of Ficoll 70 at the crowding levels up to 400 mg/ml. The base buffer for these experiments was 10 mM MES, pH 7.0, and 1 mM EDTA and the CaM concentration was 0.4 mg/ml. (E) Far-UV CD spectra of apoCaM in the presence of Ficoll 70 at crowding levels from zero to 400 mg/ml in base buffer with 100 mM KCl. (F) Thermal denaturation (CD at 222 nm) of apoCaM in the presence of Ficoll 70 at the crowding levels up to 400 mg/ml in base buffer with 100 mM KCl. The CaM concentration in all of these experiments was 0.4 mg/ml.

Figure 4. 2-D free energy landscape of apoCaM at different temperatures at [KCl]=0.5M. 2-D free energy landscape is plotted by the overlap function ($\chi$) and the asphericity ($\Delta$) in bulk (volume fraction of crowding agents, $\varphi_c=0$) in Panel A, B, and C and $\varphi_c=40\%$ in Panel D, E, and F. (A) 1.07 $k_BT/\varepsilon$, $\varphi_c=0$. (B) 1.15 $k_BT/\varepsilon$, $\varphi_c=0$. (C) 1.4 $k_BT/\varepsilon$, $\varphi_c=0$. (D) 1.07 $k_BT/\varepsilon$, $\varphi_c=40\%$. (E) 1.15 $k_BT/\varepsilon$, $\varphi_c=40\%$. (F) 1.4 $k_BT/\varepsilon$, $\varphi_c=40\%$.

Figure 5. 2-D free energy landscape of apoCaM at different ionic strengths at T = 1.15 $k_BT/\varepsilon$. 2-D free energy landscape is plotted by the overlap function ($\chi$) and the asphericity

($\Delta$) in bulk (volume fraction of crowding agents, $\varphi_c=0$) in Panel A, B, and C and $\varphi_c=40\%$ in Panel D, E, and F. (A) [KCl]=0.1M, $\varphi_c=0$. (B) [KCl]=0.2M, $\varphi_c=0$. (C) [KCl]=0.5M, $\varphi_c=0$. (D) [KCl]=0.1M, $\varphi_c=40\%$. (E) [KCl]=0.2M, $\varphi_c=40\%$. (F) [KCl]=0.5M, $\varphi_c=40\%$.

Figure 6. The helicity of CaM in different ionic strengths as a function of temperature. (A) apoCaM, $\varphi_c=0$. (B) apoCaM, $\varphi_c=40\%$. (C) holoCaM, $\varphi_c=0$. (D) A computed CD spectrum using data from simulations.

Figure 7. 2-D free energy landscape of holoCaM at different temperatures at [KCl]=0.5M. 2-D free energy landscape is plotted by the overlap function ($\chi$) and the asphericity ($\Delta$). (A) 1.07 $k_BT/\varepsilon$, $\varphi_c=0$. (B) 1.15 $k_BT/\varepsilon$, $\varphi_c=0$. (C) 1.4 $k_BT/\varepsilon$, $\varphi_c=0$.

Figure 8. Covariance matrix of contact formation of apoCaM. The covariance matrix is plotted as a function of Contact Index Pairs (see Table S3) at different ionic strengths at 1.15 $k_BT/\varepsilon$. (A) [KCl]=0.1M. (B) [KCl]=0.5M.

Figure 9. Schematic illustrations for apoCaM. Blocks represents selected helices. (A) At [KCl]=0.5M. (B) [KCl]=0.1M. Blocks from the N-terminus is colored in red and those from the C-terminus is colored in blue. The central linker is represented with a coil. A few contact pairs (a-d) are selected to illustrate the correlation between N- and C-domains. In the cartoon representations of apoCaM protein, Phe12 and Met76 are colored in orange (contact a), Ala1 and Ala147 are colored in yellow (contact b), and Phe89 and Phe141 are colored in green (contact c).

**Tables**

Table 1. Thermal Unfolding of Apo-CaM in increasing ionic strength.

| Sample (KCl mM) | model[a] | $T_m$ (°C) | | $\Delta H_m$ (kcal/mol) | | $\Delta C_p$ (cal/K*mol) | |
|---|---|---|---|---|---|---|---|
| Bulk | N-I | 24.4 | ±2.5 | 21.0 | ±4.1 | | |
| | I-U | 51.9 | ±0.6 | 41.5 | ±.09 | | |
| 50 | N-I | 41.8 | ±7.5 | 24.3 | ±2.5 | | |
| | I-U | 55.4 | ±1.4 | 43.9 | ±2.0 | | |
| 100 | N-I | 46.6 | ±3.6 | 27.2 | ±1.9 | | |
| | I-U | 61.3 | ±0.8 | 43.0 | ±1.4 | | |
| 250 | N-I | 55.2 | ±4.9 | 27.8 | ±4.1 | | |
| | I-U | 67.2 | ±1.9 | 39.0 | ±5.4 | | |
| 500 | N-I | 53.8 | ±3.8 | 31.9 | ±7.6 | | |
| | I-U | 68.2 | ±0.94 | 38.2 | ±3.5 | | |
| $Ca^{2+}$ | N-U | 94.23 | ±4.34 | 29.22 | ±1.07 | 601.04 | ±71.42 |

[a]Data of apoCaM were fit to a three-state model of unfolding where $\Delta C_p$ was fixed at zero. Data of holoCaM (+$Ca^{2+}$; from Figure 2) was fit to a two-state model of unfolding with the assumption that C-terminal of holoCaM is already unfolded at 95 °C.

Table 2. Thermal Unfolding of Apo-CaM in increasing crowder concentration.

| Sample | model[a] | $T_m$ (°C) No KCl | | $\Delta H_m$(kcal/mol) No KCl | | $T_m$(°C) 100 mM KCl | | $\Delta H_m$(kcal/mol) 100 mM KCl | |
|---|---|---|---|---|---|---|---|---|---|
| Bulk | N-I | 24.3 | ±3.1 | 21.1 | ±4.4 | 46.2 | ±3.5 | 26.5 | ±2.0 |
|  | I-U | 51.8 | ±0.8 | 41.7 | ±1.2 | 61.4 | ±0.8 | 42.9 | ±1.6 |
| 10% | N-I | 33.5 | ±2.4 | 24.7 | ±1.6 | 49.4 | ±3.3 | 28.6 | ±2.7 |
|  | I-U | 54.9 | ±0.5 | 46.0 | ±0.9 | 63.5 | ±1.0 | 43.5 | ±2.5 |
| 20% | N-I | 38.2 | ±2.7 | 26.2 | ±1.2 | 52.1 | ±3.0 | 32.1 | ±3.7 |
|  | I-U | 57.7 | ±0.7 | 45.7 | ±1.1 | 65.3 | ±1.4 | 44.5 | ±3.6 |
| 30% | N-I | 43.4 | ±3.0 | 27.6 | ±1.7 | 54.8 | ±3.1 | 34.0 | ±4.6 |
|  | I-U | 61.1 | ±0.8 | 47.7 | ±1.3 | 67.3 | ±2.0 | 45.8 | ±5.3 |
| 40% | N-I | 48.4 | ±2.1 | 29.1 | ±2.5 | 56.7 | ±2.2 | 36.3 | ±5.6 |
|  | I-U | 65.9 | ±0.9 | 49.1 | ±1.9 | 70.7 | ±3.3 | 46.9 | ±6.5 |

[a]Data were fit to a three-state model of unfolding.

Table 3. The angle, $\Theta_{ij}$, between the two helices, *i* and *j*, forming an EF hand of apoCaM (or holoCaM) in different states (M1, M2 and M3) at [KCl]=0.5M.

|  | M1 (apoCaM) | M2 (apoCaM) | M3 (holoCaM) | NMR experiment[4] |
|---|---|---|---|---|
| $\Theta_{AB}$ | 133.6±0.0 | 110.7±0.0 | 105.4±0.0 | 138±2 |
| $\Theta_{CD}$ | 129.7±0.0 | 117.9±0.0 | 96.9±0.0 | 130±3 |
| $\Theta_{EF}$ | 113.1±0.0 | 100.6±0.0 | 83.7±0.0 | 131±4 |
| $\Theta_{GH}$ | 130.8±0.0 | 117.8±0.0 | 81.2±0.0 | 133±4 |

The direction of an α-helix, *i*, is defined by a vector pointing from an averaged position of the first four residues (N) to an averaged position of the last four residues in an α-helix. $\Theta_{ij}$ of an EF hand is defined by the arccosine of the inner product of the two vectors of helices *i* and *j*. The notation of ABCDEFGH represents the eight helices of CaM from its N-terminus to C-terminus.

Fig 1

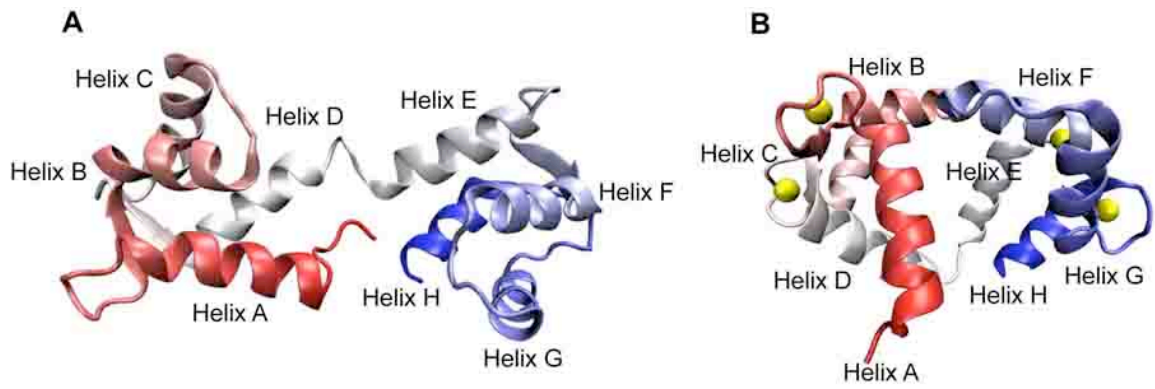

Fig 2

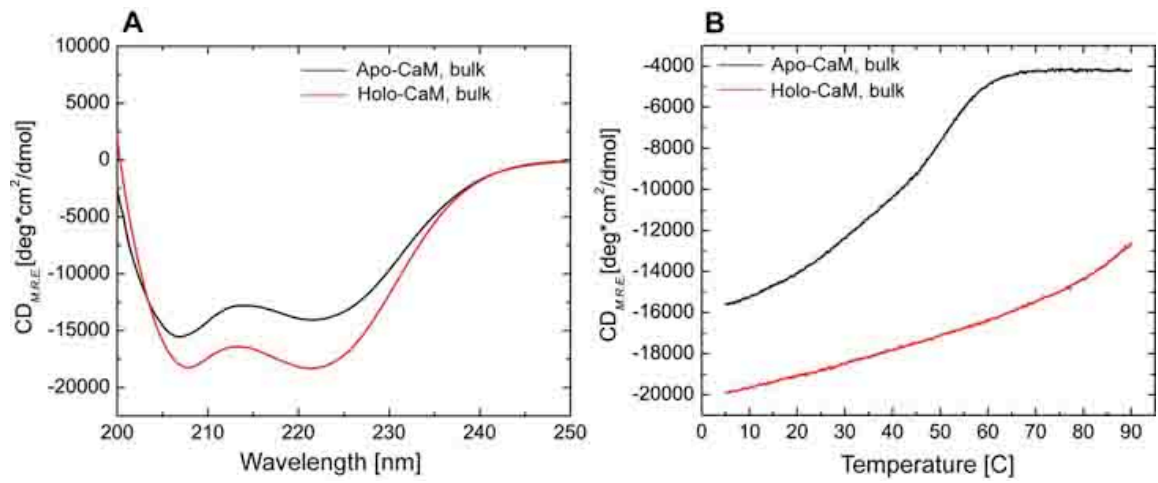

Fig 3

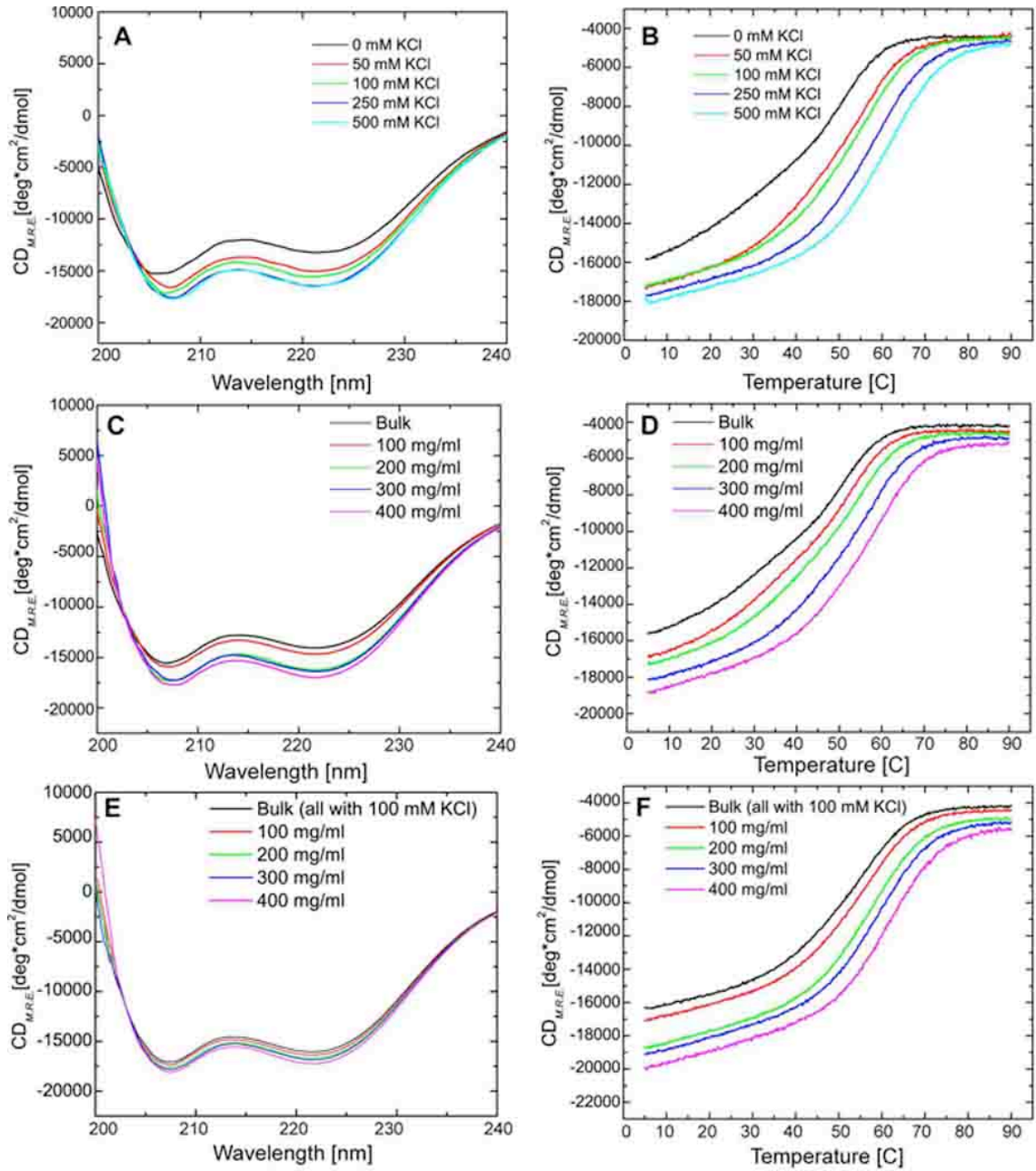

Fig 4

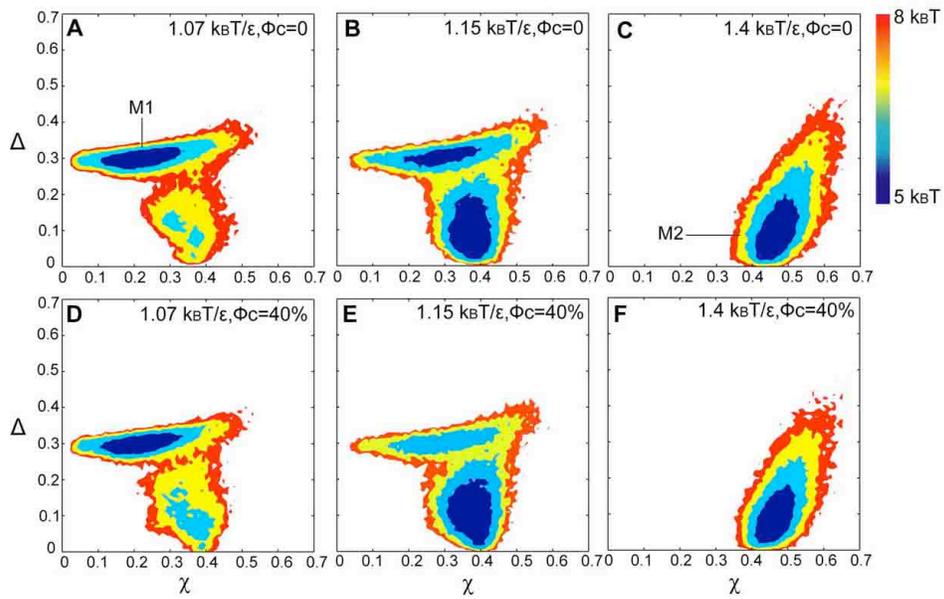

Fig 5

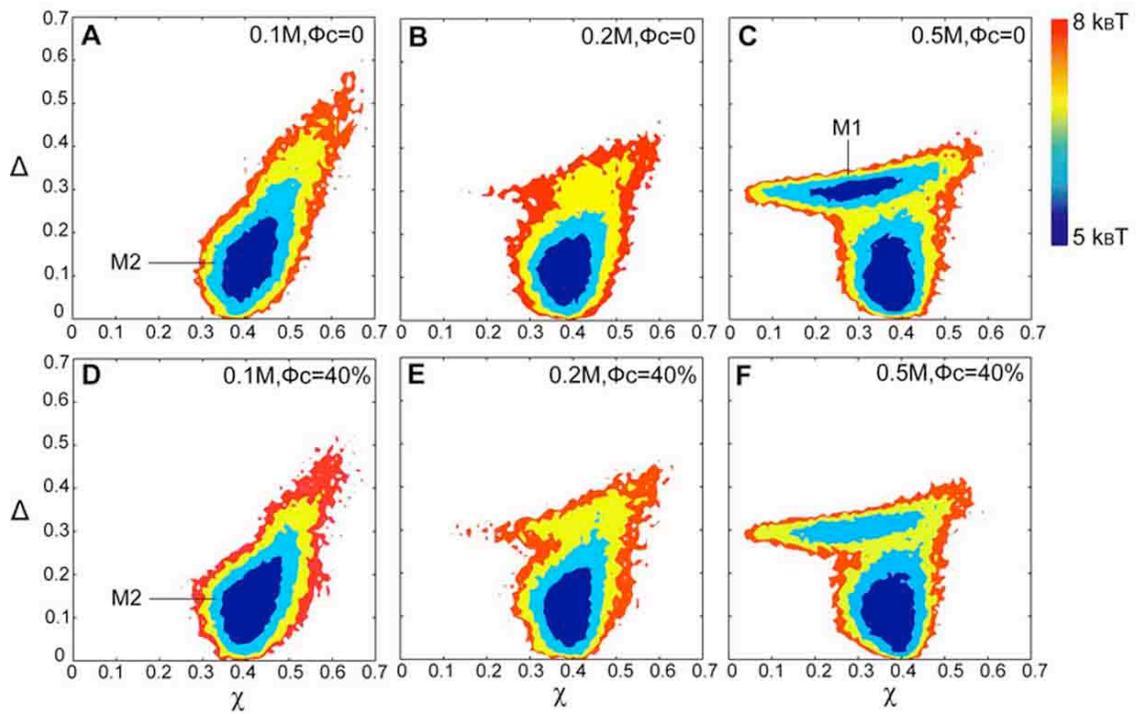

Fig 6

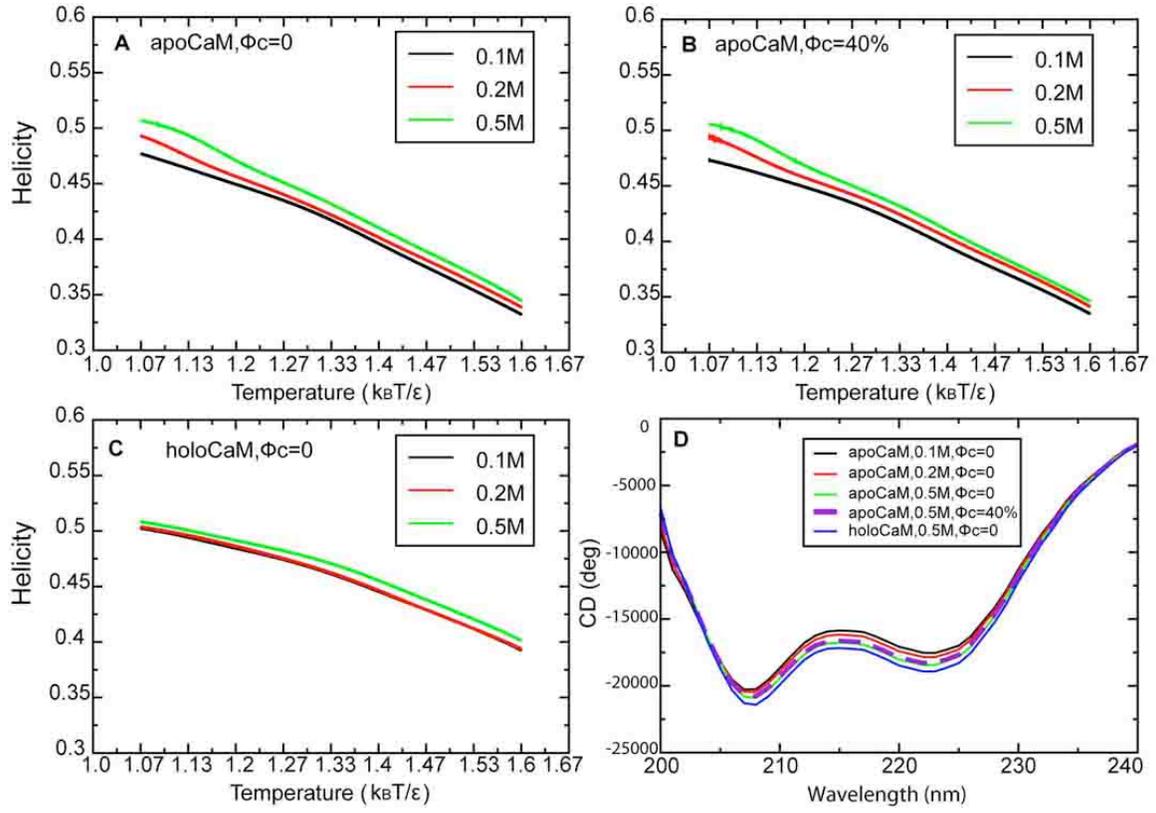

Fig 7

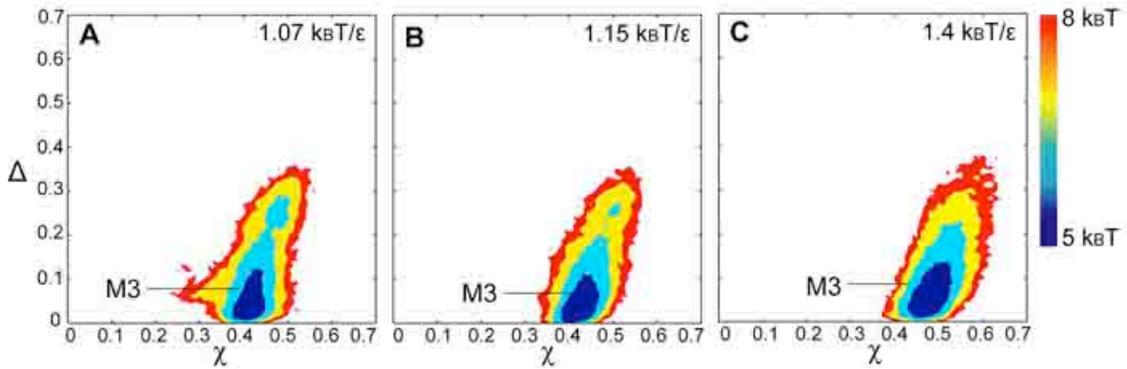

Fig 8

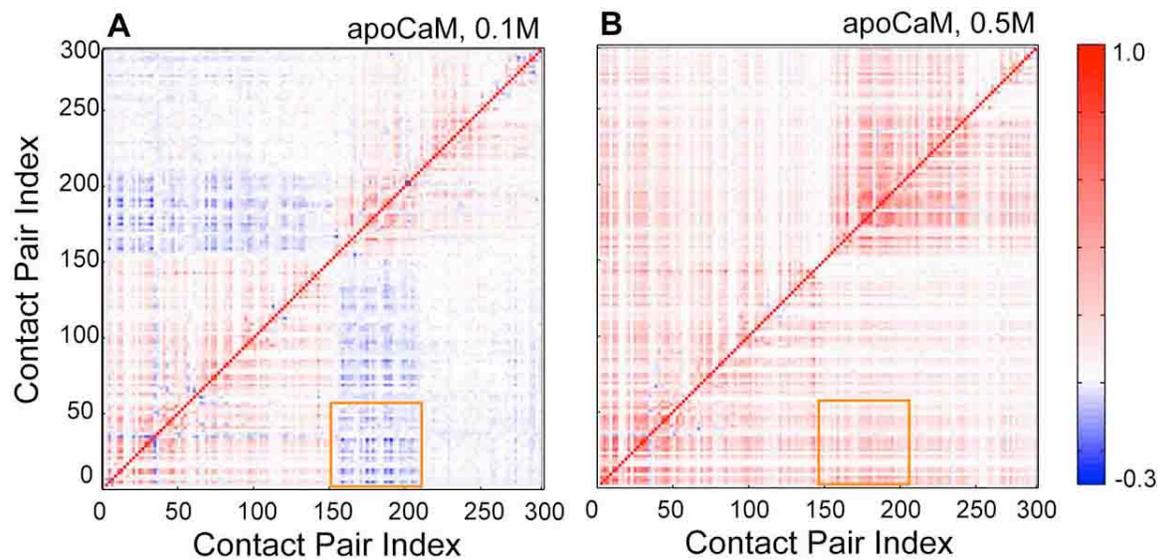

Fig 9

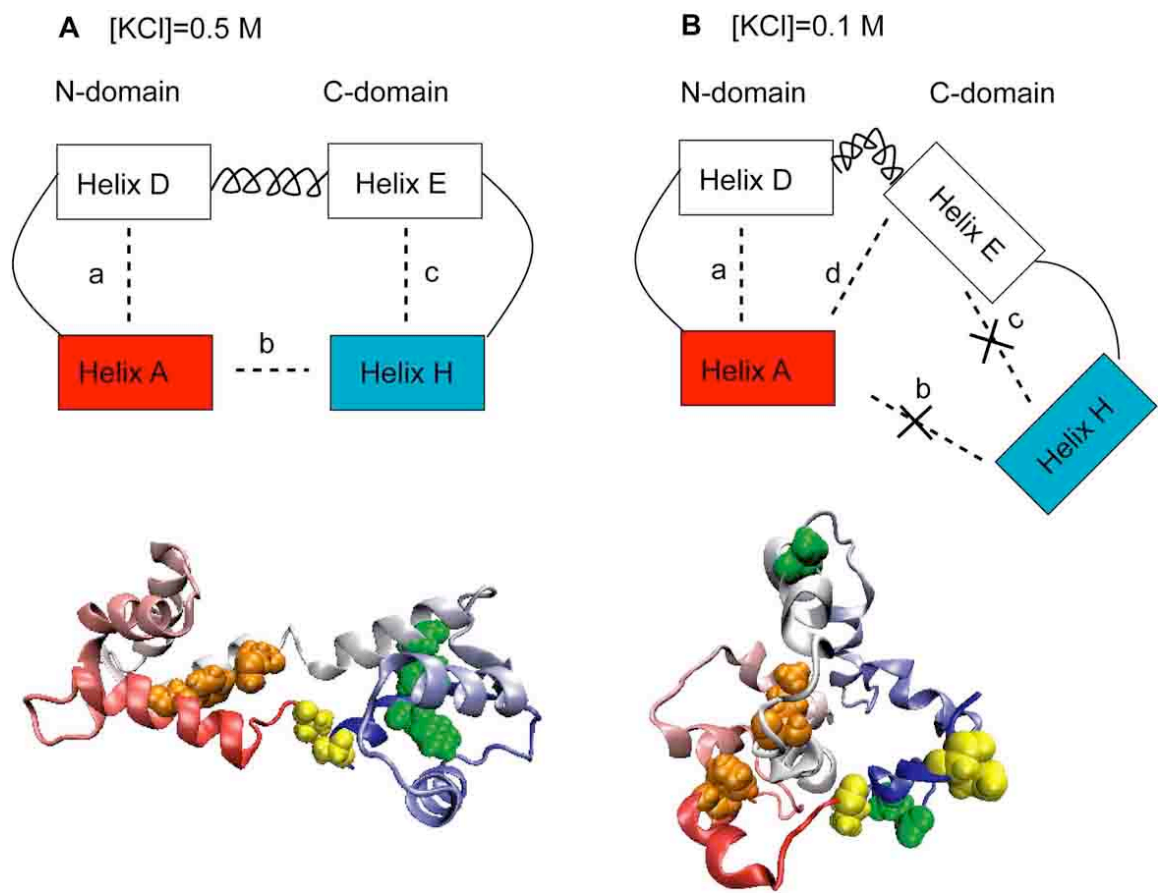

37